# THE EARLY HISTORY OF DARK MATTER

Sidney van den Bergh

Dominion Astrophysical Observatory

Herzberg Institute of Astrophysics

National Research Council of Canada

5071 West Saanich Road

Victoria, British Columbia, V8L 5L9, Canada

## ABSTRACT

The history of the discovery of dark matter in the Universe is briefly reviewed.  Special emphasis is placed on the early work by Zwicky (1933), Smith (1936), Babcock (1939) and Oort (1940).

It is not clear how these startling results must ultimately be interpreted.

Zwicky (1957)

## 1.    EARLY HISTORY

The discovery by Zwicky (1933) that visible matter accounts for only a tiny fraction of all of the mass in the Universe may turn out to have been the one of the most profound new insights produced by scientific exploration during the 20th Century.  From observations of the radial velocities of eight galaxies in the Coma cluster Zwicky found an unexpectedly large velocity dispersion $\sigma = 1019 \pm 360$ km s$^{-1}$.  [If the most deviant galaxy is rejected as a foreground object the velocity dispersion becomes $\sigma = 706 \pm 267$ km s$^{-1}$.]   It is noted in passing that Zwicky's velocity dispersion from only eight galaxies agrees well with the modern value $\sigma = 1082$ km s$^{-1}$ obtained by Colless & Dunn (1996).  Zwicky concluded from these observations that, for a velocity dispersion of 1000 km s$^{-1}$, the mean density of the Coma cluster would have to be 400 times greater than that which is derived from luminous matter.  Zwicky overestimated the mass-to-light ratio of the Coma cluster because he assumed a Hubble parameter $H_o = 558$ km s$^{-1}$ Mpc$^{-1}$.  His value for the overdensity of the Coma cluster should therefore be reduced from 400 to  ~ 50[1]. Zwicky writes (my translation):  "If this

---

[1]     It is of interest to note that Hubble's prestige was so great, that none of the early authors thought of reducing Hubble's constant as a way of lowering their mass-to-light ratios.

[overdensity] is confirmed we would arrive at the astonishing conclusion that dark mater is present [in Coma] with a much greater density than luminous matter." He continues: "From these considerations it follows that the large velocity dispersion in Coma (and in other clusters of galaxies) represents an unsolved problem." It is not yet clear what the basis was for Zwicky's claim that other clusters also exhibited a missing mass problem. Not until three years later (Smith 1936) was it found that the Virgo cluster also appears to exhibit an unexpectedly high mass. Smith made the interesting speculation that the excess mass of Virgo "represents a great mass of internebular material within the cluster". In his famous 1933 paper Zwicky also writes: "It is, of course, possible that luminous plus dark (cold) matter[1] together yield a significantly higher

---

[1] Zwicky's use of the words "<u>dunkle</u> (<u>kalte</u>) <u>Materie</u>" might be regarded as the first reference to cold dark matter, even though this expression was not used exactly in its modern sense. The term "cold dark matter", with its modern meaning, was introduced by Bond et al. (1983).

---

density...". A quarter of a century later Kahn & Woltjer (1959) pointed out that M 31 and the Galaxy were moving towards each other, so that they must have completed most of a (very elongated) orbit around each other during a Hubble time. If M 31 and the Galaxy started to move apart 15 Gyr these authors found that the mass of the Local Group had to be $\geq 1.8 \times 10^{12} \, M_\odot$. Assuming that the combined mass of the Andromeda galaxy and the Milky Way system was 0.5 x $10^{12}$ M$_\odot$, Kahn & Woltjer concluded that most of the mass of the Local Group existed in some invisible form. They opined that it was most likely that this missing mass was in the form of very hot (5 x $10^5$ K) gas. From a historical perspective, it is interesting to note that Kahn & Woltjer did not seem to have been aware of the earlier papers by Zwicky (1933) and Smith (1936) on missing mass in clusters of galaxies.

## 2.     THE DARK AGES

Six years after Zwicky's paper Babcock (1939) obtained long-slit spectra of the Andromeda galaxy, which showed that the outer regions of M 31 were rotating with an unexpectedly high velocity, indicating either (1) a high outer mass-to-light ratio or (2) strong dust absorption. Babcock wrote: "[T]he great range in the calculated ratio of mass to luminosity in proceeding outward from the nucleus suggests that absorption plays a very important rôle in the outer portions of the spiral, or, perhaps, that new dynamical considerations are required, which will permit of a smaller relative mass in the outer parts". Subsequently Babcock's optical rotation curve, and that of Rubin & Ford (1970), was extended to even larger radii by Roberts and Whitehurst (1975) using 21-cm line observations that reached a radial distance of ~ 30 kpc. These observations clearly showed that the rotation curve of M 31 did not exhibit a Keplerian drop-off. In fact, its rotational velocity remained constant over radial distances of 16 - 30 kpc. These observations indicated that the mass in the outer regions of the Andromeda galaxy increased with galactocentric distance, even though the optical luminosity of M 31 did not. From these observations Roberts & Whitehurst concluded that the mass-to-light ratio had to be $\geq$ 200 in the outermost regions of the Andromeda galaxy. It is interesting to note that neither Babcock, nor Roberts & Whitehurst, cited the 1933 paper by Zwicky. In other words, no connection was made between the missing mass in the outer region of a spiral and the missing mass in rich clusters such as Coma (Zwicky 1933), and Virgo (Smith 1936). Roberts Whitehurst suggested that the very high mass-to-light ratio that they observed in the outer regions of M 31 might be attainable by postulating the presence of a vast population of dM5 stars in the outer reaches of the Andromeda galaxy. Regarding the flat outer rotation curves of galaxies Roberts (1999, private communication) recalls that this result "was, at best, received with skepticism in many colloquia and meeting presentations."

From a historical perspective the paper by Roberts and Whitehurst was important because it, together with papers on the stability of galactic disks by Ostriker & Peebles (1973), and on the apparent increase of galaxy mass with increasing radius (Ostriker, Peebles & Yahil 1973), first convinced the majority of astronomers that missing mass existed. Ostriker & Peebles concluded that bar-like instabilities in galaxy disks could be prevented by a massive spherical (halo) component. They wrote that "[T]he halo masses of our Galaxy and of other spiral galaxies exterior to the observed disks may be extremely large."  However, this conclusion was later contested by Kalnajs (1987), who concluded that massive bulges are more efficient at stabilizing disks, than are halos.  "[I]n the end it may well be that a massive halo is an important stabilizing influence on most galactic disks"  (Binney & Tremaine 1987).

Observations obtained during the last quarter century have strongly supported the conclusions by Ostriker, Peebles & Yahil that (1) the mass-to-light ratios of galaxies grow larger with increasing radius, and (2) that this missing mass is large enough to be cosmologically significant.  The inference that observations of the inner regions of galaxies underestimate their total mass had already been anticipated by Zwicky  (1937), who wrote:  "Present estimates of the masses of nebulae are based on observations of the <u>luminosities</u> and <u>internal rotations</u> of nebulae.  It is shown that both these methods are unreliable; that from the observed luminosities of extragalactic systems only lower limits for the values of their masses can be obtained".  It is noted in passing that Zwicky (1937) also points out that gravitational lensing might provide useful information on the total masses of galaxies.

With his unusually fine nose for "smelling" the presence of interesting astronomical problems Oort (1940) studied the rotation and surface brightness of the edge-on S0 galaxy NGC 3115.  He found that "[T]he distribution of mass in this system appears to bear almost no relation to that of light." He concluded that M/L ~ 250 in the outer regions of NGC 3115. However, this value is reduced by almost an order of magnitude if modern distances to this galaxy are adopted. Oort ended his paper by writing that "There cannot be any doubt that an extension of the measures of rotation to greater distances from the nucleus would be of exceptional interest." Again no connection was made between the missing mass in this S0 galaxy, and the Zwicky/Smith missing mass problem in rich clusters of galaxies. Finally X-ray observations of early-type galaxies (Forman, Jones & Tucker 1985), which provide a unique tracer for the gravitational potential in the outer regions of these objects, confirmed that they must be embedded in massive coronae.

No good detective story is complete without at least one false clue. Oort (1960, 1965) believed that he had found some dynamical evidence for the presence of missing mass in the disk of the Galaxy. If true, this would have indicated that some of the dark matter was dissipative in nature. However, late in his life, Jan Oort told me that the existence of missing mass in the Galactic plane was never one of his most firmly held scientific beliefs. Recent observations, that have been reviewed by Tinney (1999), show that brown dwarfs cannot make a significant contribution to the density of the Galactic disk near the Sun.

A third line of evidence (see Table 1) for the possible existence of dark

Place Table 1 here

matter was provided by a statistical analysis of the separations, and velocity

differences, between the members of pairs of galaxies.  The data in the table are

seen to be somewhat ambiguous, but not inconsistent with the notion that

significant amounts of non-luminous matter might be associated with galaxy

pairs.  The idea that we are looking at the same missing mass phenomenon in

galaxies, in pairs, and in clusters was first discussed in detail at the Santa Barbara

Conference on the Stability of Systems of Galaxies (Neyman, Page & Scott

1961).  In his introduction to this conference, and after excluding various

alternatives, Ambartsumian (1961) concluded that:  "Thus there is only one

natural assumption left relating to the clusters cited above - they have positive

total energies."  If this conclusion were correct, then rich clusters of galaxies

would disintegrate, and scatter their galaxian content into the field, on a time-

scale that is short compared to the age of the Universe.  However, van den Bergh

(1962) pointed out that this hypothesis had to be incorrect because such a large fraction of all early-type galaxies are presently cluster members.  He therefore concluded:  "That such a large fraction of all galaxies are at present members of clusters suggests that most clusters are stable over periods comparable to their ages."  If clusters of galaxies are stable, then we are stuck with the high mass-to-light ratios, that are calculated from application of the virial theorem.  Einasto,

Kaasik & Saar (1974) first pointed out that the gas in such rich clusters, which had been discovered from its X-radiation, did not have a large enough mass to stabilize these clusters. For an excellent review on dark matter during the "Dark Ages" the reader is referred to Ashman (1992).

## 3.    RECENT HISTORY

By 1975 the majority of astronomers had become convinced that missing mass existed in cosmologically significant amounts. However, it was not yet clear whether this mass was in the form of late M dwarfs, brown dwarfs, white dwarfs, black holes, very hot gas, or in some, as yet unsuspected, form. Rees (1977) concluded that: "There are other possibilities of more exotic character - for instance the idea of neutrinos with small (few ev) rest mass has been taken surprisingly seriously by some authors". In other words it was not yet clear in 1977 that a paradigm shift (Kuhn 1962) would be required to interpret the new observations that seemed to support the ubiquitousness of missing matter in the Universe. Alternatively, it has also been speculated (Milgrom & Bekenstein 1987) that such a paradigm shift might not be required if Newton's laws break down at small accelerations. The idea that neutrinos (hot dark matter) could provide the missing matter was downplayed by White, Davis & Frenk (1984) who concluded that "[T]he properties of the neutrino aggregates expected in a neutrino-dominated universe are incompatible with observations irrespective of the efficiency with  which they form galaxies." The very high dark matter densities in dwarf spheroidal galaxies, in conjunction with the Pauli exclusion principle, also place severe constraints on neutrino dark matter models (Tremaine & Gunn 1979). The idea that the missing mass might be in the form of "cold dark matter", which dominates current speculation on this subject, appears to have been introduced by Bond et al. (1983). Current "best buy" models of the Universe (Roos & Harun-or- Rashio 1999, Turner 1999) suggest that cold dark matter accounts for $30 \pm 10\%$ of the closure density of the Universe, compared to only $4 \pm 1\%$ in the form of baryonic matter. If such models are correct then more than 2/3 of the closure density is in the form of vacuum energy, or in even more exotic states (Wang et al. 1999).

It is a pleasure to thank Dick Bond, Don Osterbrock, Jim Peebles, Mort Roberts, and Simon White for sharing some of their historical reminiscences with me. I also wish to thank the editors of PASP for inviting me to expand a short talk on the early history of dark matter, that I gave at the Dominion Astrophysical Observatory last year, into a paper for these Publications.

**TABLE 1**

**Mass-to-light ratios in binary systems [1]**

| Galaxy type | M/L(Page 1960) | M/L(van den Bergh 1961) |
|---|---|---|
| E + E | 66 ± 27 | 35 ± 14 |
| E + S | 34 ± 20 | 18 ± 16 |
| S+S and S+Ir | 0.2 ± 0.2 | 3 ± 3 |

[1]     $H_o = 70$ km s$^{-1}$ Mpc$^{-1}$ assumed.